\documentclass[sigconf]{acmart}

\usepackage{booktabs} 
\usepackage{todonotes}
\usepackage{microtype}
\usepackage{xspace}
\usepackage{balance}
\usepackage{paralist}

\renewcommand{\paragraph}[1]{\vspace{0.08em}\noindent {\bf #1}}

\AtBeginDocument{%
  \providecommand\BibTeX{{%
    \normalfont B\kern-0.5em{\scshape i\kern-0.25em b}\kern-0.8em\TeX}}}

\copyrightyear{2020}
\acmYear{2020}
\setcopyright{acmcopyright}
\acmConference[ITiCSE '20]{Proceedings of the 2020 ACM Conference on Innovation and Technology in Computer Science Education}{June 15--19, 2020}{Trondheim, Norway}
\acmBooktitle{Proceedings of the 2020 ACM Conference on Innovation and Technology in Computer Science Education (ITiCSE '20), June 15--19, 2020, Trondheim, Norway}
\acmPrice{15.00}
\acmDOI{10.1145/3341525.3387389}
\acmISBN{978-1-4503-6874-2/20/06}

\newtheorem{definition}{Definition}

\newcommand{\summary}[2]{
        \vspace{0.4em}
        \noindent
        \colorbox{gray!20}{%
            \parbox{.97\linewidth}{%
                    \textbf{\textsf{Summary (\textit{#1})}}
                #2
            }%
        }%
}%

\begin{document}
\fancyhead{}

\title{Common Bugs in Scratch Programs}


\settopmatter{authorsperrow=5}

\author{\mbox{Christoph Frädrich}}
\email{fraedric@fim.uni-passau.de}
\affiliation{%
  \institution{University of Passau}
  \city{Passau}
  \country{Germany}
}
\author{\mbox{Florian Obermüller}}
\email{obermuel@fim.uni-passau.de}
\affiliation{%
  \institution{University of Passau}
  \city{Passau}
  \country{Germany}
}
\author{Nina Körber}
\email{koerber@fim.uni-passau.de}
\affiliation{%
  \institution{University of Passau}
  \city{Passau}
  \country{Germany}
}

\author{Ute Heuer}
\email{ute.heuer@uni-passau.de}
\affiliation{%
  \institution{University of Passau}
  \city{Passau}
  \country{Germany}
}

\author{Gordon Fraser}
\email{gordon.fraser@uni-passau.de}
\affiliation{%
  \institution{University of Passau}
  \city{Passau}
  \country{Germany}
}
\renewcommand{\shortauthors}{Frädrich, et al.}

\begin{abstract}

Bugs in \scratch{} programs can spoil the fun and inhibit learning success.
Many common bugs are the result of recurring patterns of bad code.
In this paper we present a collection of common code patterns that typically hint at bugs in \scratch{} programs, and the \litterbox tool which can automatically detect them.
We empirically evaluate how frequently these patterns occur, and how severe their consequences usually are.
While fixing bugs inevitably is part of learning, the possibility to identify the bugs automatically provides
the potential to support learners.

\end{abstract}



\newcommand{\litterbox}{\textsc{LitterBox}\xspace}
\newcommand{\scratch}{\textsc{Scratch}\xspace}
\newcommand{\drscratch}{\textsc{Dr. Scratch}\xspace}
\newcommand{\hairball}{\textsc{Hairball}\xspace}
\newcommand{\qualityhound}{\textsc{Quality Hound}\xspace}
\newcommand{\findbugs}{\textsc{FindBugs}\xspace}

\newcommand{\numtotal}{135,164\xspace}
\newcommand{\numnoremix}{74,830\xspace}
\newcommand{\numexcluded}{60,334\xspace}
\newcommand{\numbuggyprojects}{33,655\xspace}
\newcommand{\numbugs}{109,951\xspace}

\newcommand{\numbuggyproc}{1,162\xspace}
\newcommand{\numprocprojects}{10,576\xspace}
\newcommand{\numpenprojects}{5,303\xspace}
\newcommand{\numbuggypen}{1,218\xspace}
\newcommand{\numbugpattern}{25\xspace}

\newcommand{\numclassified}{250\xspace}
\newcommand{\numfailure}{70\xspace}
\newcommand{\numnofail}{96\xspace}
\newcommand{\numnoexec}{52\xspace}
\newcommand{\numfalsepositives}{32\xspace}


\maketitle

\section{Introduction}

Block-based programming languages like \scratch~\cite{maloney2010}
are hugely successful at introducing and engaging young learners with
the concepts of programming, and once children are hooked they use
their imagination to produce complex and intricate games, stories,
and other types of programs.
Often these programs do not work immediately because of \emph{bugs},
i.e., mistakes in the assembled blocks. While finding and fixing such
bugs (i.e., \emph{debugging} programs) is an essential skill and an
important aspect of learning, bugs can nevertheless be the source of
endless frustration, discouraging learners and potentially
inhibiting their learning success.

\begin{figure}[t]
	\centering
	\includegraphics[width=0.4\columnwidth]{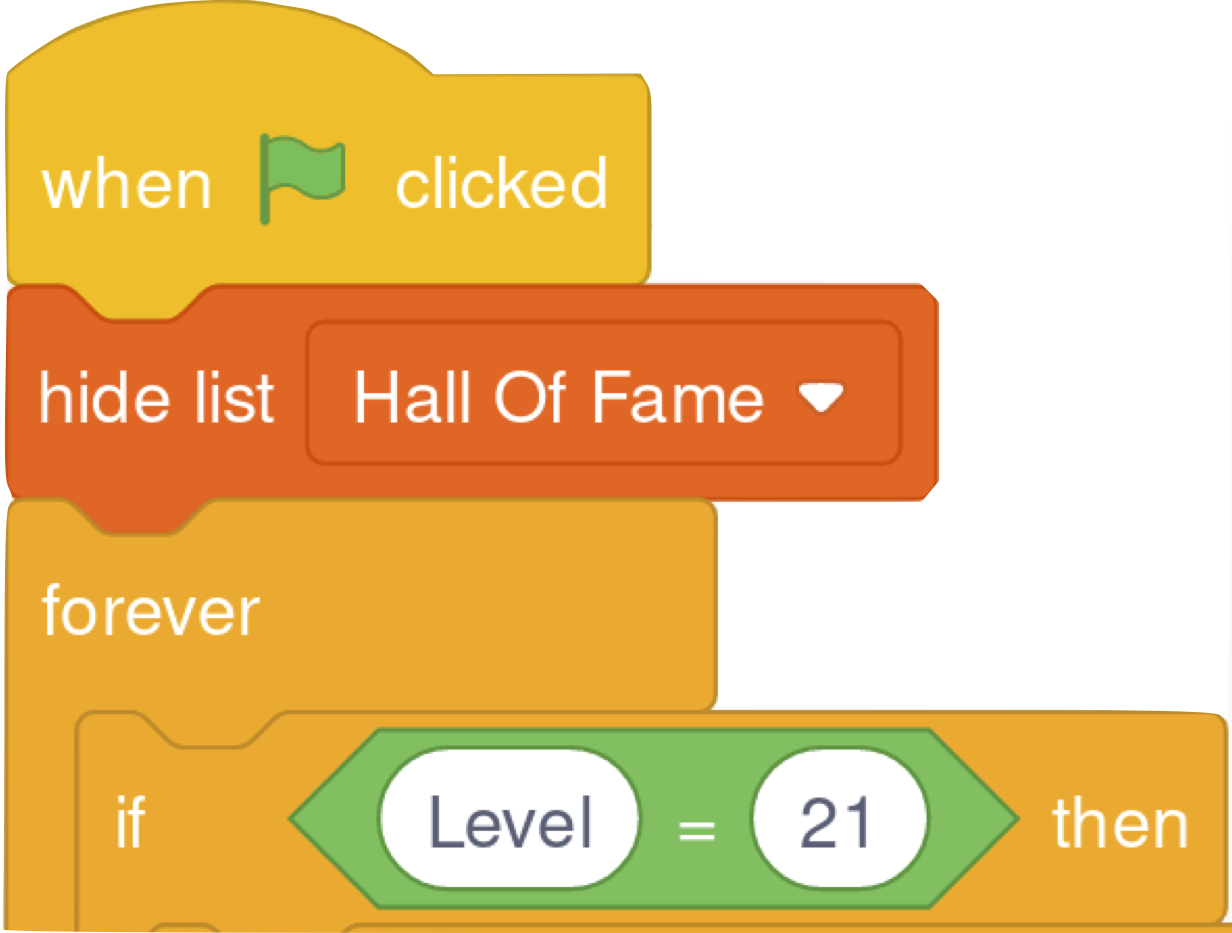}
	\vspace{-1em}
	\caption{\label{fig:example1}Common \scratch bug, taken from a publicly shared project: The comparison of the literals ``Level'' and ``21'' in the if-condition will never be true. Instead of ``Level'', the user should have used the variable with the same name.}
	\vspace{-1em}
\end{figure}

Bugs can occur for different reasons, for example when programming
concepts are not well understood, or simply because it is challenging
to produce correctly working programs, regardless of whether one
writes code in \scratch or any other programming language. Often,
code that works initially is written in a clumsy and confusing
way (commonly referred to as code that \emph{smells}, such as long scripts or duplicated code~\cite{hermans2016a,techapalokul2017a}). This causes bugs to
be introduced later, when changing and extending the existing code.
%
While there is no end to the creativity with which learners produce
bugs, many of these bugs are the result of similar misunderstandings,
and manifest in similar, reoccurring \emph{patterns} of bugs. For example, consider Figure~\ref{fig:example1}: The comparison of two literals frequently happens if variables are not fully comprehended or if code is incomplete.
Identifying such patterns offers the potential to support learners.

In this paper, we present a catalogue of \numbugpattern \emph{bug patterns}
based on common misunderstandings and programming errors in \scratch. We
introduce \litterbox, an automated program analysis tool which can
find occurrences of these bug patterns in \scratch programs. Given a
\scratch project ID, \litterbox retrieves and parses the source code
of the project, and reports all instances of bug patterns identified.
To investigate how common the bug patterns are in practice, we
applied \litterbox to a random sample of \numnoremix \scratch projects.
Since the occurrence of a bug pattern does not guarantee that a
program is broken, we further investigated the typical severity of
each of the bug patterns by manually checking how they affect real
\scratch programs.

Our investigations show that a common misunderstanding lies in the usage of
key-event handlers as the best way of reacting to user input (resulting in
\emph{Stuttering Movement}), and that synchronising scripts with
messages or backdrops (e.g., \emph{Missing Backdrop Switch}) can be error
prone. Some bug patterns like \emph{Stuttering Movement} almost always break
programs, while some bugs are more subtle, like the \emph{Position Equals Check}, where positions are checked using exact equality instead of more permissive relations, resulting in fragile and misbehaving programs.
In either case, however, an instance of a bug pattern is a sign of a problem.
The possibility to identify bugs automatically can help
learners with the inevitable debugging part of learning, and in producing working \scratch programs.


\section{Background}
\label{sec:background}

In this paper we investigate bug patterns in \scratch{}~\cite{maloney2010} programs.
\scratch is a widely used visual block-based programming language that aims at making programming more accessible for novices. It favours exploration over recall, implements visualisation of grammar rules and visually distinguishes different categories of statements and expressions~\cite{bau2017}. 
%
%
%
The block-based nature of \scratch prevents syntactical errors in the program code, but it is nevertheless possible to produce \scratch code that is problematic or plain wrong.
Bad programming practices often lead to code that uses inefficient, confusing, or awkward constructions to achieve its result. In programming, characteristics of source code that represent bad quality and are indicative of deeper problems are commonly referred to as \emph{code smells}:

\begin{definition}[Code Smell]
	A code smell is a code idiom that increases the likelihood of
bugs in a program \cite{fowler1999}.
\end{definition}

\begin{table}[tb]
	\caption{\label{tab:smells}Common code smells for \scratch.}
\begin{tabular}{ll}
	\toprule
		Code Smell & Reference \\
	\midrule
Broad variable scope & \cite{techapaloku2017b} \\
Complex animation& \cite{boe2013} \\
Dead code & \cite{techapaloku2017b,hermans2016b,boe2013} \\
Duplicate code & \cite{techapaloku2017b,hermans2016b,boe2013} \\
Duplicate string & \cite{techapaloku2017b} \\
Empty script  & \cite{techapaloku2017b,hermans2016b} \\
Empty sprite & \cite{hermans2016b} \\
Feature envy & \cite{techapaloku2017b,hermans2016b} \\
Inappropriate intimacy & \cite{techapaloku2017b,hermans2016b} \\
Long script & \cite{techapaloku2017b,hermans2016b} \\
Many parameters & \cite{hermans2016b} \\
Middle man & \cite{techapaloku2017b,hermans2016b} \\
Uncommunicative name & \cite{techapaloku2017b,boe2013} \\
Unused variable & \cite{techapaloku2017b,hermans2016b} \\
	\bottomrule
\end{tabular}
\end{table}

It has been established that certain code smells such as long methods
or duplicated code are prevalent in \scratch
programs~\cite{hermans2016a,techapalokul2017a}. Categories of code
smells in \scratch are mostly derived by mapping smells from other
programming languages~\cite{hermans2016a} but also by defining new
smells that exist only in block based languages~\cite{moreno2014}.
There is also evidence that code smells can have a negative effect on
the learning progress of novice programmers~\cite{hermans2016b}. 
Table~\ref{tab:smells} summarises the code smells that have been defined for \scratch.

Smelly code is not wrong per se; the program likely still works
correctly. However, smells typically decrease understandability and
increase the chances of programmers using and extending the code
wrongly in the future, thus introducing \emph{bugs}.
The term \emph{bug} is often ambiguously used; when talking about bugs in this paper we refer to \emph{defects} as defined in the IEEE Standard
Classification of Software Anomalies~\cite{ieee1044}:

\begin{definition}[Defect]
	A defect is a weakness in code that prevents software from
meeting its specification and needs to be fixed~\cite{ieee1044}.
\end{definition}

Defects in programs can stem from misconceptions affecting the programmer's problem solving process. This is a serious challenge for novices even in \scratch\cite{swidan2018}.
A defect represents functionally incorrect code, but the consequences of the defect are not necessarily visible to the user of the program. Only under certain conditions does a defect result in an observable deviation from the correct and expected behaviour. For example, the defect often has to be executed in a particular way in order to cause the program to misbehave, thus causing a \emph{failure}:

\begin{definition}[Failure]\label{def:failure}
   A failure represents the termination of the ability of a program to perform its required function. The execution of a defect may result in a failure.
\end{definition}

In practice, failures often result in program crashes, exploitable
security holes, data-loss, or other severe consequences. In \scratch
programs a failure represents a change in the output on the
stage, such as erroneous movement of a sprite, wrong backdrops being
shown, lack of reaction to a user input, and it might also result in
the execution stopping unexpectedly or generally decrease the
usability of the program (e.g., by slowing the program execution).
On one hand, failures can be incredibly frustrating to learners.
On the other hand, productive failure is seen as an emerging innovation in pedagogic practice and is considered a promising approach in computer science education as well~\cite{falkner2019}.

Failures are typically found through \emph{testing}, i.e., running
the program and checking the result against a specification
(potentially
automatically~\cite{stahlbauer2019testing,johnson2016itch}). Given a
failure, \emph{debugging} describes the activities of (1) identifying
the underlying defect, and (2) producing an appropriate fix for the
defect. When debugging programmers may gain insights on the programming task, into its pitfalls and proper solutions. Related pedagogic interventions should facilitate and strengthen these insights. Therefore meaningful teaching and training activities can encourage and guide reflection and communication about critical thinking involved in the debugging process.~\cite{deliema2019,kafai2019}. However, not all defects
easily manifest in failures, potentially making them hard to spot.
Failing to identify defects reduces chances for learners to practice
fixing. Failing to fix defects, programs may never work.

In order to support learners in identifying defects, we therefore aim
to automatically identify defects in \scratch programs. For this, we
define and identify \emph{bug patterns}. A bug pattern is a code
idiom that is likely to be a defect~\cite{hovemeyer2004}.

\begin{definition}[Bug Pattern]
   A bug pattern in \scratch is a composition of blocks typical of defective code, or a common erroneous deviation of a correct code idiom.
\end{definition}

The difference between a code smell and a bug pattern is that a code
smell reports negative attributes of the code (e.g., long, confusing,
duplicate, unused code), whereas bug patterns report specific
combinations of blocks that are indicative of defects. Although
previous work on code quality in \scratch focused on code
smells, some of these are actually bug patterns, in particular the \emph{undefined block}
smell~\cite{techapaloku2017b} as well as \hairball's checks for (1)
matching broadcast and receive blocks, (2) synchronisation of say and
sound blocks, and (3) proper initialisation of attributes and
variables. Note that the occurrence of a bug pattern does not
guarantee the existence of a defect (i.e., there may be \emph{false
positives}):

\begin{definition}[False Positive]
 	Non-defective code that matches a bug pattern constitutes a false positive.
\end{definition}

False positives are common for static program analysis tools~\cite{hovemeyer2004}, and may occur for any approximate analysis. In particular, for bug patterns false positives may result if it is not possible to describe a combination of blocks that \emph{precisely} distinguishes all defective from non-defective cases.
Bug patterns are commonly used in professional
programming, where tools like \findbugs\cite{hovemeyer2004} implement checks for
catalogues of common bug patterns for the Java language.
In this paper, we aim to
define a catalogue of bug patterns for \scratch.

\section{Bug Patterns in \scratch}
\label{sec:patterns}

In this section we describe bug patterns for \scratch{} programs.
These patterns mostly originate from our experiences of teaching and building an analysis infrastructure for \scratch{} programs, and some are 
specialised versions of
code smells~\cite{boe2013, hermans2016a}.
We organise patterns in three categories: (1) Bugs that are effectively syntax errors, as a compiler would detect them in a text-based language; (2) bugs that can occur in any programming language; and (3) bug patterns that are specific to~\scratch.

\subsection{Syntax Errors}

\paragraph{Ambiguous Custom Block Signature:}
\scratch does not enforce unique names for custom blocks.
Two custom blocks with the same name can only be distinguished if they have a
different number or order of parameters.
When two blocks have the same name and parameter order, no matter which call block is used, the program will always execute the custom block which was defined earlier.

\paragraph{Ambiguous Parameter Name:}
The parameter names in custom blocks do not have to be unique.
When two parameters have the same name, no matter the type or which one is used inside the custom block, it will always be evaluated as the last parameter.

\paragraph{Call Without Definition:}
When a custom block is called without being defined nothing happens.
This can occur in two different situations:
1) Earlier releases of \scratch 3 allowed removal of a custom block definition even when the custom block is still in use.
2) A script using a call to a custom block can be dragged and copied to another sprite, probably no custom block with the same signature as the call exists here and thus the call has no definition.

\paragraph{Expression As Touchable Or Colour:}
This happens when inside a block that expects a colour or sprite as parameter (e.g., \textit{set pen color to} or
\textit{touching mouse-pointer?}) a reporter block, or an expression with a string or number value is used.

\paragraph{Missing Termination Condition:}
The \textit{repeat until} blocks require a stopping condition.
If the condition is missing, the result is an infinite loop.
This will then prevent the execution of blocks following the loop in the script.

\paragraph{Orphaned Parameter:}
When custom blocks are created the user can define parameters, which can then be used in the body of the custom block.
However, the block definition can be altered, including removal of parameters even if they are in use.
Any instances of deleted parameters are retained, and then evaluated with the default value for the type of parameter in the underlying JavaScript execution of the \scratch virtual machine (i.e., $0$ or empty string), since they are never initialised.

\paragraph{Parameter Out Of Scope:}
The parameters of a custom block can be used anywhere inside the sprite that defines the custom block.
However, they will never be initialised outside the custom block, and will always have the default value.

\subsection{General Bugs}

\paragraph{Comparing Literals:}
Reporter blocks can be used to evaluate the truth value of certain expressions.
Not only is it possible to compare literals to variables or the results of other reporter blocks, literals can also be compared to literals (e.g. Fig. \ref{fig:example1}).
Since this will lead to the same result in each execution this construct is unnecessary and it can obscure that certain blocks will never or always be executed.

\paragraph{Custom Block With Forever:}
If a custom block contains a \textit{forever} loop and the custom block is used in the middle of another script, that other script will never be able to complete its execution. The \textit{forever} loop in the custom block cannot be left, resulting in the calling script never being able to proceed.

\paragraph{Custom Block With Termination:}
If a custom block contains a \textit{Stop all} or \textit{Delete this clone} and the custom block is called in the middle of another script, the script will never reach the blocks following the call.

\paragraph{Endless Recursion:}
If a custom block calls itself inside its body and has no condition to stop the recursion, it will run for an indefinite amount of time.

\paragraph{Forever Inside Loop:}
If two loops are nested and the inner loop is a \textit{forever} loop, the inner loop will never terminate. Thus any statements preceeding the inner loop are only executed once (e.g. Fig. \ref{fig:example-loop}). Furthermore, any further statements following the outer loop can also never be reached.

\paragraph{Message Never Received:}
This pattern is a specialised version of \textit{unmatched broadcast and receive blocks}~\cite{boe2013}. It occurs when there are blocks to send messages, but the \textit{When I receive message} event handler is missing. Since no handler reacts to this event, the message stays unnoticed.

\paragraph{Message Never Sent:}
This pattern is a specialised version of \textit{unmatched broadcast and receive blocks}~\cite{boe2013}. When there are blocks to receive messages but the corresponding \textit{broadcast message} block is missing, that script will never be executed.

\paragraph{Missing Clone Call:}
If the \textit{When I start as a clone} event handler is used to start a script, but the sprite is never cloned, the event will never be triggered and the script is dead.

\paragraph{Missing Clone Initialisation:} When a sprite creates a
clone of itself but has no scripts started by \textit{When I start as
a clone} or \textit{When this sprite clicked} events, clones will not
perform any actions. The clones remain frozen until they
are deleted by \textit{delete this clone} blocks or the program is
restarted.

\paragraph{Missing Loop Sensing:}
If a script is supposed to execute actions conditionally when an event occurs, this is often done by continuously checking for the event inside a \textit{forever} or \textit{until} loop. If the loop is missing, the occurrence of the event is only checked once and thus likely missed.

\paragraph{No Working Scripts:}
The empty script smell (cf. Lazy Class \cite{hermans2016a}) occurs if an event handler has no other blocks attached to it.
The dead code smell \cite{hermans2016a} occurs when a script has no event handler and can never be executed automatically.
If both smells occur simultaneously without any other scripts in a sprite we consider it a bug, since the script should likely consist of the event handler attached to the dead code.

\paragraph{Position Equals Check:}
\scratch uses floating point values to store positions and calculate distances to other sprites or the mouse-pointer. Since two floating point values might never match exactly, using exact comparisons of these values as guards in conditional statements or loops such as \textit{until}/\textit{wait until} is prone to failure.

\paragraph{Recursive Cloning:}
Scripts starting with a \textit{When I start as a clone} event handler that contain a \textit{create clone of myself} block may result in an infinite recursion.

\subsection{Scratch-specific Bugs}

\paragraph{Missing Backdrop Switch:}
If the \textit{When backdrop switches to} event handler is used to start a script and the backdrop never switches to the selected one,  the script is never executed. This does not apply to programs including at least one of the switch options \textit{next}, \textit{previous} or \textit{random}.

\paragraph{Missing Erase All:}
If a sprite uses a \textit{pen down} block but never an \textit{erase all} block, then all drawings from a previous execution might remain, making it impossible to get a blank background without reloading the \scratch{} project.

\paragraph{Missing Pen Down:}
Scripts of a sprite using a \textit{pen up} block but never a \textit{pen down} are likely wrong since either the sprite is supposed to draw
something and does not, or the pen up may intefere with later additions of pen down blocks.

\paragraph{Missing Pen Up:}
A sprite that uses \textit{pen down} blocks but never a \textit{pen up} may draw right away, when the project is restarted.
This might not be intended.

\paragraph{Stuttering Movement:}
A common way to move sprites in response to keyboard input is to use the specific event handler \textit{When key pressed} followed by a \textit{move steps}, \textit{change x by} or \textit{change y by} statement.
Compared to the alternative to use a \textit{forever} loop with a conditional containing a \textit{key pressed?} expression, the first approach results in noticeably slower reaction and stuttering movement of the sprite moved.

\section{Evaluation}

We aim to answer the following research questions:
\begin{compactdesc}
	\item \textbf{RQ1}: How common are the bug patterns?

	\item \textbf{RQ2}: How severe are the consequences of the defects?
\end{compactdesc}
RQ1 aims to give an overview on how many projects are affected by each bug pattern.
By answering RQ2 our goal is to understand how often bug patterns cause failures. 

\subsection{Experimental Setup}

\paragraph{Analysis tool: } We implemented the bug patterns listed in Section~\ref{sec:patterns} in our own Java-based
static analysis tool \litterbox. For the analysis \litterbox first constructs an abstract syntax tree (AST). Each bug
pattern finder is implemented as a visitor of the AST, and application to a \scratch program reports all instances of the bug pattern found in the AST.

\paragraph{Dataset: } Since \litterbox parses \scratch projects in the new data format introduced with \scratch version 3, we could not use existing datasets (e.g.,~\cite{aivaloglou2017dataset}).
Therefore we created a new dataset of real \scratch programs by downloading the most recent programs from the \scratch platform over a period of 3 weeks.
For this we first downloaded the project IDs of the most recent projects via the \scratch REST
API\footnote{\url{https://github.com/LLK/scratch-rest-api/wiki}, last accessed 17.1.2020}
which \litterbox then used to directly download the JSON project file from the project host\footnote{\url{https://projects.scratch.mit.edu}, last accessed 17.1.2020}.
On average between 5,000 and 10,000 new programs were created each day, resulting in \numtotal projects for our dataset.
A replication package containing all data and software needed to reproduce the results can be found at \url{https://github.com/se2p/artifact-iticse2020}.


\paragraph{RQ1: } To answer RQ1 we applied \litterbox on the data set.
Since it is common to start a new \scratch{} project by remixing an existing one it may also happen that bug patterns are inherited from the original projects. Counting the same bug patterns in remixed code multiple times would potentially skew our analysis of bug pattern frequency.
Therefore we excluded all projects that were remixes. This led to the exclusion of \numexcluded projects with \numnoremix remaining for the analysis.
We consider the cumulative number of bug patterns found as well as the number of projects that contain at least one instance of each pattern. We further consider the weighted method count (WMC, i.e., sum of cyclomatic complexity of all scripts) to quantify the average complexity of projects containing bug patterns.

\paragraph{RQ2: } To answer RQ2 we sampled 10 affected projects for each bug pattern from our data set, excluding remixes. 
We then manually classified whether the occurrence of the pattern results in a failure based on
Definition~\ref{def:failure}.
Based on this definition we only counted occurrences of a pattern as a failure if it has a noticeable effect for the user.
Thus, the classification required understanding each project and its intended behaviour, as well as executing and playing with the program in order to derive a verdict.
In case a bug pattern did not yield a failure, we further classified whether this was a) due to the bug pattern being contained in code that was not executed at all, b) the occurrence of a false positive or c) for other reasons.
Each project was independently classified by two authors.
In case of disagreements the bug pattern instances were discussed. If no consensus was reached, at least one
more author was consulted.

\paragraph{Threats To Validity: }
%
To avoid skewing the numbers, 
we exclude remixes in our analysis.
Our dataset consists of only publicly shared projects;
it is conceivable that bug patterns might occur more frequently in unfinished projects not yet shared.
To avoid bias we randomised the selection of projects.
For the manual classification (RQ2), each project was independently classified by at least two authors of this paper.

\begin{table}[t!]
    \caption{Number of projects affected by each pattern and number of pattern instances found in total.}%
    \label{tab:patterns_across_projects}
	\resizebox{\columnwidth}{!}{%
        \begin{tabular}{@{}l r r r@{}}%
        \toprule%
        Pattern                         & \#Projects & \#Patterns & AVGWMC \\ %
        \midrule%
        Ambiguous CB Signature          & 92         & 317    & 138.74 \\ %
        Ambiguous Parameter Name        & 25         & 402    & 39.28  \\ %
        Call Without Definition         & 164        & 569    & 288.2  \\ %
        Comparing Literals              & 1,999       & 4,939   & 242.95 \\ %
        Custom Block With Forever       & 176        & 574    & 313.58 \\ %
        Custom Block With Termination   & 263        & 985    & 243.6  \\ %
        Endless Recursion               & 109        & 386    & 141.54 \\ %
        Expression As Touchable Or Color& 365        & 1,076   & 217.79 \\ %
        Forever Inside Loop             & 2,158       & 9,617   & 192.48 \\ %
        Message Never Received          & 5,933       & 12,479  & 163.95 \\ %
        Message Never Sent              & 4,781       & 24,933  & 193.99 \\ %
        Missing Backdrop Switch         & 903        & 4,282   & 154.47 \\ %
        Missing Clone Call              & 625        & 1,522   & 356.33 \\ %
        Missing Clone Initialization    & 1,691       & 5,862   & 185.71 \\ %
        Missing Erase All               & 164        & 245    & 73.12  \\ %
        Missing Loop Sensing            & 3,282       & 8,372   & 130.54 \\ %
        Missing Pen Down                & 164        & 199    & 203.17 \\ %
        Missing Pen Up                  & 988        & 1,931   & 47.98  \\ %
        Missing Termination             & 709        & 1,564   & 257.43 \\ %
        No Working Scripts              & 549        & 730    & 86.61  \\ %
        Orphaned Parameter              & 88         & 175    & 308.19 \\ %
        Parameter Out Of Scope          & 461        & 1,493   & 400.15 \\ %
        Position Equals Check           & 1,472       & 4,725   & 206.81 \\ %
        Recursive Cloning               & 953        & 3,318   & 267.74 \\ %
        Stuttering Movement             & 5,541       & 19,256  & 34.02  \\ \bottomrule%
        \end{tabular}
}
\end{table}

\subsection{RQ1: How Common are the Bug Patterns?}

In the evaluation of our data set without remixes we found instances of all the bug patterns defined in Section~\ref{sec:patterns}.
Table~\ref{tab:patterns_across_projects} shows how many projects are affected by each bug pattern, how often each pattern occurs in total and what the average weighted method count of affected projects is.
Of the \numnoremix projects \numbuggyprojects contained at least one bug pattern, resulting in \numbugs bug patterns in total.
The most common bug patterns in terms of affected projects are \emph{Stuttering Movement} (5,541), \emph{Message Never Received} (5,933), and \emph{Message Never Sent} (4,781).
The least common are \emph{Ambiguous Parameter Name} (25), \emph{Orphaned Parameter} (88) and \emph{Ambiguous Custom Block Signature}~(92).

Projects with \emph{Stuttering Movement} arguably usually still provide the expected functionality, but the pattern
causes the projects to be almost unusable, as is well known to the \scratch{} community and frequently discussed in
forums and educational material\footnote{\url{https://en.wikibooks.org/wiki/Scratch/Lessons/Movement\#Smooth\_Movement},
last accessed 17.1.2020}.
The pattern is nevertheless often used as a means to introduce beginners as it allows controlling sprites without the
need for conditional statements. The very low average WMC of 34.02 for projects with \emph{Stuttering Movement} confirms that they are simple and small.

Bug patterns related to the pen feature are also used in simple projects (e.g., average WMC of 47.98 for \emph{Missing Pen Up}), which often have little functionality besides drawing basic patterns. However, only  \numpenprojects projects in our datasets use pen blocks at all, which is likely because pen blocks are an optional extension since \scratch~3.0.

Bug patterns related to custom blocks (e.g., \emph{Ambiguous
Parameter Name} or \emph{Ambiguous Custom Block Signature}) are less
common. The main reason for this is that custom blocks are an
advanced programming concept in \scratch, to which students are
usually introduced later. As a result, out of all the projects only
\numprocprojects used custom blocks. The average WMC confirms that
these projects tend to be more complex (e.g., 313.58 for \emph{Custom Block With Forever}).


Besides bugs related to custom blocks, the \emph{Missing Clone Call} bug pattern stands out because it is contained in more complex projects (average WMC of 356.33). 
This is most likely because it is easier to miss that a sprite is not cloned in big projects.
A notable example was a project where the \emph{When Green Flag clicked} event triggered the same behaviour as the \emph{When I start as clone} event,
%
suggesting that cloning is a topic that is not easy to understand right away.
Message-related bugs may result from missing refactoring support: When removing one part of a synchronised communication, \scratch does not warn the user that the other part remains.

\summary{RQ1}{Bugs frequently appear in \scratch{} programs regardless of their complexity, 
 and often follow a similar structure which we can use to automatically identify them.}

\begin{figure*}[t!]
    \centering
    \includegraphics[width=\textwidth]{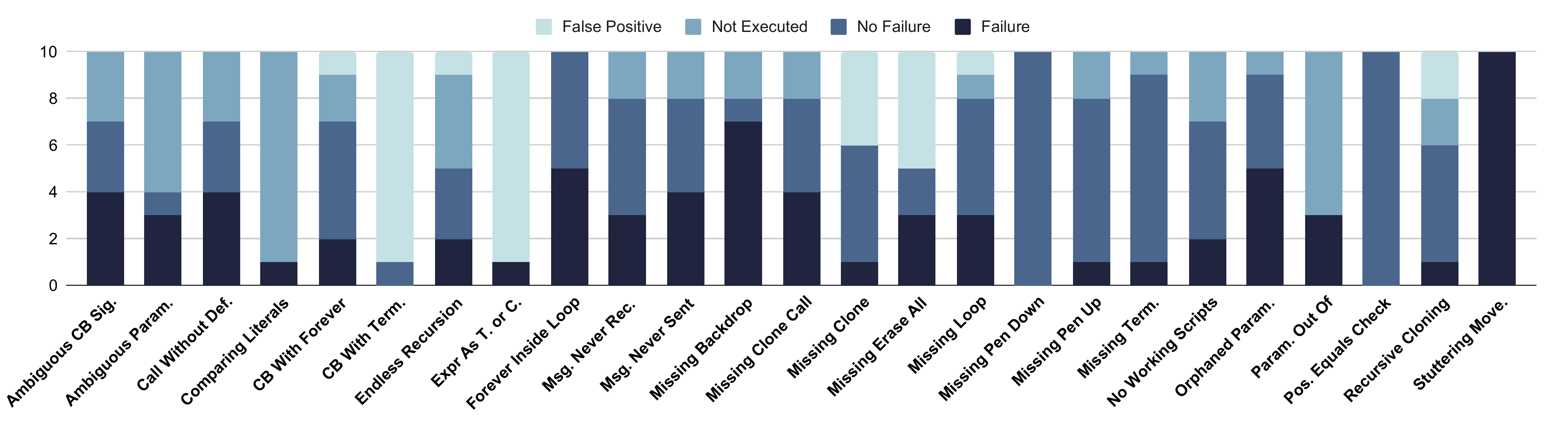}
	\vspace{-2em}
    \caption{Results of the manual classification of \scratch projects.}%
    \label{fig:classification}
\end{figure*}

\begin{figure}[t!]
	\centering
	\includegraphics[width=0.4\columnwidth]{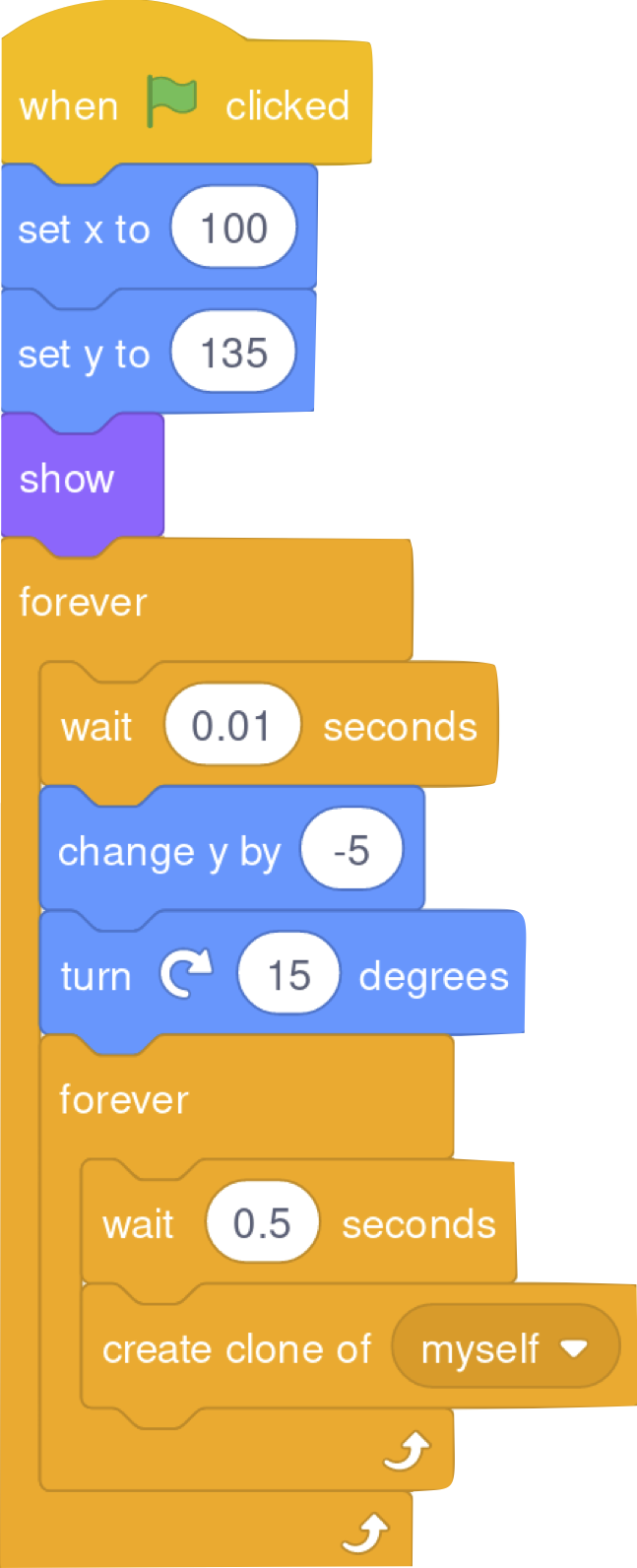}
	\caption{\label{fig:example-loop}Example of a \emph{Forever Inside Loop} bug pattern, taken from a publicly shared project: The blocks in the outer forever-loop will only be executed once, since the execution will not leave the inner forever-loop.}
\end{figure}

\subsection{RQ2: How Severe are the Consequences of the Defects?}

Figure~\ref{fig:classification} shows the results of the manual
classification for each bug pattern. Out of the \numclassified
projects we inspected, \numfailure instances of bugs manifested into
failures. Of the remaining bug patterns, \numnofail were defects without visible impact, \numnoexec did not result in failures because
the defective code was never executed, and \numfalsepositives bugs
were identified as false positives.

The most severe failures cause their project to stop working completely (i.e., no reactions to user inputs and sprites not performing any actions),
and sometimes even slow down the web browser to the point where it becomes unusable.
For example, this happened in a project with an instance of \emph{Forever Inside Loop} with multiple nested loops. Figure~\ref{fig:example-loop} shows an example of this bug pattern taken from one of the \scratch projects in our analysis: Since the inner forever-loop is never left, all clones are created at the same location and the same orientation, breaking the intended graphical effect.

The defects most frequently causing failures are \emph{Stuttering Movement} (10 of 10 executed) and \emph{Missing Backdrop Switch} (7
of 8 executed).
The frequent occurrence of \emph{Stuttering Movement} is not surprising as this
pattern is often even taught to students as a ``correct'' or
``first'' approach to animating sprites.
Notably, the latter can be created without modifying
code: Most affected projects had scripts that waited for a
backdrop that no longer existed. This leads us to believe that
this bug often
happens by accident or that
 students often do not clean up their
projects. 

Some failures are subtle: For example, some projects have multiple
scripts triggered by different messages, resulting in similar
behaviour (e.g., switching to different costumes of a sprite). In
this scenario a \emph{Message Never Sent} bug may be difficult to notice as it rarely manifests as a failure.

Some defects do not cause a failure when the program is
run for the first time, but only in later executions. For example, the
%
 \emph{Missing Pen Up} pattern may only be noticed when the program is reset
and the sprite clutters the screen.

There are, however, also multiple reasons why a bug may exist in a project and not lead to a failure.
The most common and simple reason is that the code containing the bug is never executed, as was the case in \numnoexec projects.
A reason why a bug is never executed can be as simple as it being located in unreachable code
(e.g., inside an if with a condition that always evaluates to false).
For example, the \emph{Comparing
Literals} pattern caused only one actual failure, whereas in the
other 9 cases the defect was located in disconnected blocks that were not executed,
suggesting that the authors of \scratch{} programs may be somewhat
aware of this problem.

Sometimes programs also simply ignore the wrong behaviour such as
custom blocks with \emph{Ambiguous Parameter Names} that do not use
the parameters at all.
Sometimes the sprites with the bug are affected in their behaviour
(e.g., position, movement, or size), but do not result in a failure because they are hidden.
An example are hidden sprites, with custom blocks that contain a forever loop that never stops.
The \emph{Position Equals Check} caused no failures because the sprites always arrived at the checked positions;
moving the sprite only one pixel away broke the program in most cases.
In other cases the bug did not lead to failures because the author of the program worked around the
bug; for example, a \emph{Missing Pen Up} may not lead to any
noticeable problems if the program erases all drawings multiple times.

As anticipated, we also found several cases (\numfalsepositives) of false
positives where the creator of the project
deliberately used the mechanism we consider as a bug and also took
measures to prevent a failure. For example, the \emph{Custom Block
With Termination} pattern may be correct if the termination statement
is contained in conditional code. As our static analysis is an
overapproximation it currently cannot correctly detect this case.
Similarly, \emph{Recursive Cloning} produces false positives when a
clone takes care of deleting itself.

\summary{RQ2}{When defective code is executed, it frequently results in failures, but dead code and redundant expressions often prevent visible impact.}

\section{Conclusions}

Bug patterns help to identify bugs and provide a common vocabulary to
people talking about these bugs. To this end we introduced and
empirically evaluated a new catalogue of \numbugpattern bug patterns
in \scratch. Our evaluation found occurrences for each of the bug
patterns, which shows that the concept of bug patterns can be
successfully transferred to \scratch.

In this paper we focused on the actual bug patterns, but an important next step will be to study the effects of these bugs on the learning success of novice programmers, as well as guidelines for instructors on how to teach students about these patterns.
Detection of bug patterns might enable learners to engage in debugging processes more easily since the cognitive load of localising a potential bug can be reduced.
This fosters discussion of successful and unsuccessful debugging activities and helps students reflect on underlying missing concepts or misconceptions. In the future it would be interesting to investigate whether bug patterns result from misconceptions in programming as some of these seem to be a symptom of those \cite{swidan2018, sorva2018}.
In this light, productive failure seems to be a promising approach that we will pursue and investigate further in the future.

We also plan to further improve our \litterbox tool with additional
bug patterns and lower false positive rates. \litterbox is available at:
\url{https://github.com/se2p/LitterBox}
%

\vspace{-0.3em}
\section*{Acknowledgements}
This work is supported by DFG project
FR 2955/3-1  ``Testing, Debugging, and Repairing
Blocks-based Programs''. We would like to thank Florian Sulzmaier and Andreas Stahlbauer for their contributions to \litterbox.


\balance

\bibliographystyle{ACM-Reference-Format}
\bibliography{references}


\begin{thebibliography}{19}


\ifx \showCODEN    \undefined \def \showCODEN     #1{\unskip}     \fi
\ifx \showDOI      \undefined \def \showDOI       #1{#1}\fi
\ifx \showISBNx    \undefined \def \showISBNx     #1{\unskip}     \fi
\ifx \showISBNxiii \undefined \def \showISBNxiii  #1{\unskip}     \fi
\ifx \showISSN     \undefined \def \showISSN      #1{\unskip}     \fi
\ifx \showLCCN     \undefined \def \showLCCN      #1{\unskip}     \fi
\ifx \shownote     \undefined \def \shownote      #1{#1}          \fi
\ifx \showarticletitle \undefined \def \showarticletitle #1{#1}   \fi
\ifx \showURL      \undefined \def \showURL       {\relax}        \fi
\providecommand\bibfield[2]{#2}
\providecommand\bibinfo[2]{#2}
\providecommand\natexlab[1]{#1}
\providecommand\showeprint[2][]{arXiv:#2}

\bibitem[\protect\citeauthoryear{??}{iee}{2010}]%
        {ieee1044}
 \bibinfo{year}{2010}\natexlab{}.
\newblock \showarticletitle{IEEE Standard Classification for Software
  Anomalies}.
\newblock \bibinfo{journal}{\emph{IEEE Std 1044-2009 (Revision of IEEE Std
  1044-1993)}} (\bibinfo{date}{Jan} \bibinfo{year}{2010}),
  \bibinfo{pages}{1--23}.
\newblock
\showISSN{null}
\urldef\tempurl%
\url{https://doi.org/10.1109/IEEESTD.2010.5399061}
\showDOI{\tempurl}


\bibitem[\protect\citeauthoryear{Aivaloglou, Hermans, Moreno-Le{\'o}n, and
  Robles}{Aivaloglou et~al\mbox{.}}{2017}]%
        {aivaloglou2017dataset}
\bibfield{author}{\bibinfo{person}{Efthimia Aivaloglou},
  \bibinfo{person}{Felienne Hermans}, \bibinfo{person}{Jes{\'u}s
  Moreno-Le{\'o}n}, {and} \bibinfo{person}{Gregorio Robles}.}
  \bibinfo{year}{2017}\natexlab{}.
\newblock \showarticletitle{A dataset of scratch programs: scraped, shaped and
  scored}. In \bibinfo{booktitle}{\emph{Proceedings of the 14th International
  Conference on Mining Software Repositories}}. IEEE,
  \bibinfo{pages}{511--514}.
\newblock


\bibitem[\protect\citeauthoryear{Bau, Gray, Kelleher, Sheldon, and Turbak}{Bau
  et~al\mbox{.}}{2017}]%
        {bau2017}
\bibfield{author}{\bibinfo{person}{David Bau}, \bibinfo{person}{Jeff Gray},
  \bibinfo{person}{Caitlin Kelleher}, \bibinfo{person}{Josh Sheldon}, {and}
  \bibinfo{person}{Franklyn Turbak}.} \bibinfo{year}{2017}\natexlab{}.
\newblock \showarticletitle{Learnable Programming: Blocks and Beyond}.
\newblock \bibinfo{journal}{\emph{Commun. ACM}} \bibinfo{volume}{60},
  \bibinfo{number}{6} (\bibinfo{date}{May} \bibinfo{year}{2017}),
  \bibinfo{pages}{72–80}.
\newblock
\showISSN{0001-0782}
\urldef\tempurl%
\url{https://doi.org/10.1145/3015455}
\showDOI{\tempurl}


\bibitem[\protect\citeauthoryear{Boe, Hill, Len, Dreschler, Conrad, and
  Franklin}{Boe et~al\mbox{.}}{2013}]%
        {boe2013}
\bibfield{author}{\bibinfo{person}{Bryce Boe}, \bibinfo{person}{Charlotte
  Hill}, \bibinfo{person}{Michelle Len}, \bibinfo{person}{Greg Dreschler},
  \bibinfo{person}{Phillip Conrad}, {and} \bibinfo{person}{Diana Franklin}.}
  \bibinfo{year}{2013}\natexlab{}.
\newblock \showarticletitle{Hairball: Lint-inspired static analysis of scratch
  projects}.
\newblock \bibinfo{journal}{\emph{SIGCSE 2013 - Proceedings of the 44th ACM
  Technical Symposium on Computer Science Education}},
  \bibinfo{pages}{215--220}.
\newblock
\urldef\tempurl%
\url{https://doi.org/10.1145/2445196.2445265}
\showDOI{\tempurl}


\bibitem[\protect\citeauthoryear{Deliema, Dahn, Flood, Asuncion, Abrahamson,
  Enyedy, and Steen}{Deliema et~al\mbox{.}}{2019}]%
        {deliema2019}
\bibfield{author}{\bibinfo{person}{David Deliema}, \bibinfo{person}{Maggie
  Dahn}, \bibinfo{person}{Virginia Flood}, \bibinfo{person}{Ana Asuncion},
  \bibinfo{person}{Dor Abrahamson}, \bibinfo{person}{Noel Enyedy}, {and}
  \bibinfo{person}{Francis Steen}.} \bibinfo{year}{2019}\natexlab{}.
\newblock \bibinfo{booktitle}{\emph{Debugging as a Context for Fostering
  Reflection on Critical Thinking and Emotion}}.
\newblock \bibinfo{pages}{209--228}.
\newblock
\showISBNx{9780429323058}
\urldef\tempurl%
\url{https://doi.org/10.4324/9780429323058-13}
\showDOI{\tempurl}


\bibitem[\protect\citeauthoryear{Falkner and Sheard}{Falkner and
  Sheard}{2019}]%
        {falkner2019}
\bibfield{author}{\bibinfo{person}{Katrina Falkner} {and} \bibinfo{person}{Judy
  Sheard}.} \bibinfo{year}{2019}\natexlab{}.
\newblock \bibinfo{booktitle}{\emph{Pedagogic Approaches}}.
\newblock \bibinfo{publisher}{Cambridge University Press},
  \bibinfo{pages}{445–480}.
\newblock
\urldef\tempurl%
\url{https://doi.org/10.1017/9781108654555.016}
\showDOI{\tempurl}


\bibitem[\protect\citeauthoryear{Fowler}{Fowler}{1999}]%
        {fowler1999}
\bibfield{author}{\bibinfo{person}{Martin Fowler}.}
  \bibinfo{year}{1999}\natexlab{}.
\newblock \bibinfo{booktitle}{\emph{Refactoring: Improving the Design of
  Existing Code}}.
\newblock \bibinfo{publisher}{Addison-Wesley}, \bibinfo{address}{Boston, MA,
  USA}.
\newblock
\showISBNx{0-201-48567-2}


\bibitem[\protect\citeauthoryear{Hermans and Aivaloglou}{Hermans and
  Aivaloglou}{2016}]%
        {hermans2016b}
\bibfield{author}{\bibinfo{person}{Felienne Hermans} {and}
  \bibinfo{person}{Efthimia Aivaloglou}.} \bibinfo{year}{2016}\natexlab{}.
\newblock \showarticletitle{Do code smells hamper novice programming? A
  controlled experiment on Scratch programs}. In \bibinfo{booktitle}{\emph{2016
  IEEE 24th International Conference on Program Comprehension (ICPC)}}.
  \bibinfo{pages}{1--10}.
\newblock
\showISSN{null}
\urldef\tempurl%
\url{https://doi.org/10.1109/ICPC.2016.7503706}
\showDOI{\tempurl}


\bibitem[\protect\citeauthoryear{Hermans, Stolee, and Hoepelman}{Hermans
  et~al\mbox{.}}{2016}]%
        {hermans2016a}
\bibfield{author}{\bibinfo{person}{Felienne Hermans},
  \bibinfo{person}{Kathryn~T. Stolee}, {and} \bibinfo{person}{David
  Hoepelman}.} \bibinfo{year}{2016}\natexlab{}.
\newblock \showarticletitle{Smells in Block-Based Programming Languages}. In
  \bibinfo{booktitle}{\emph{2016 {{IEEE Symposium}} on {{Visual Languages}} and
  {{Human}}-{{Centric Computing}} ({{VL}}/{{HCC}})}} (2016-09).
  \bibinfo{publisher}{{IEEE}}, \bibinfo{pages}{68--72}.
\newblock
\showISBNx{978-1-5090-0252-8}
\urldef\tempurl%
\url{https://doi.org/10.1109/VLHCC.2016.7739666}
\showDOI{\tempurl}


\bibitem[\protect\citeauthoryear{Hovemeyer and Pugh}{Hovemeyer and
  Pugh}{2004}]%
        {hovemeyer2004}
\bibfield{author}{\bibinfo{person}{David Hovemeyer} {and}
  \bibinfo{person}{William Pugh}.} \bibinfo{year}{2004}\natexlab{}.
\newblock \showarticletitle{Finding Bugs is Easy}.
\newblock \bibinfo{journal}{\emph{SIGPLAN Not.}} \bibinfo{volume}{39},
  \bibinfo{number}{12} (\bibinfo{date}{Dec.} \bibinfo{year}{2004}),
  \bibinfo{pages}{92–106}.
\newblock
\showISSN{0362-1340}
\urldef\tempurl%
\url{https://doi.org/10.1145/1052883.1052895}
\showDOI{\tempurl}


\bibitem[\protect\citeauthoryear{Johnson}{Johnson}{2016}]%
        {johnson2016itch}
\bibfield{author}{\bibinfo{person}{David~E Johnson}.}
  \bibinfo{year}{2016}\natexlab{}.
\newblock \showarticletitle{ITCH: Individual Testing of Computer Homework for
  Scratch Assignments}. In \bibinfo{booktitle}{\emph{Proceedings of the 47th
  ACM Technical Symposium on Computing Science Education}}. ACM,
  \bibinfo{pages}{223--227}.
\newblock


\bibitem[\protect\citeauthoryear{Kafai, DeLiema, Fields, Lewandowski, and
  Lewis}{Kafai et~al\mbox{.}}{2019}]%
        {kafai2019}
\bibfield{author}{\bibinfo{person}{Yasmin~B. Kafai}, \bibinfo{person}{David
  DeLiema}, \bibinfo{person}{Deborah~A. Fields}, \bibinfo{person}{Gary
  Lewandowski}, {and} \bibinfo{person}{Colleen Lewis}.}
  \bibinfo{year}{2019}\natexlab{}.
\newblock \showarticletitle{Rethinking Debugging as Productive Failure for CS
  Education}. In \bibinfo{booktitle}{\emph{Proceedings of the 50th ACM
  Technical Symposium on Computer Science Education}}
  \emph{(\bibinfo{series}{SIGCSE ’19})}. \bibinfo{publisher}{Association for
  Computing Machinery}, \bibinfo{address}{New York, NY, USA},
  \bibinfo{pages}{169–170}.
\newblock
\showISBNx{9781450358903}
\urldef\tempurl%
\url{https://doi.org/10.1145/3287324.3287333}
\showDOI{\tempurl}


\bibitem[\protect\citeauthoryear{Maloney, Resnick, Rusk, Silverman, and
  Eastmond}{Maloney et~al\mbox{.}}{2010}]%
        {maloney2010}
\bibfield{author}{\bibinfo{person}{John Maloney}, \bibinfo{person}{Mitchel
  Resnick}, \bibinfo{person}{Natalie Rusk}, \bibinfo{person}{Brian Silverman},
  {and} \bibinfo{person}{Evelyn Eastmond}.} \bibinfo{year}{2010}\natexlab{}.
\newblock \showarticletitle{The Scratch Programming Language and Environment}.
\newblock \bibinfo{journal}{\emph{ACM Transactions on Computing Education
  (TOCE)}}  \bibinfo{volume}{10} (\bibinfo{date}{11} \bibinfo{year}{2010}),
  \bibinfo{pages}{16}.
\newblock
\urldef\tempurl%
\url{https://doi.org/10.1145/1868358.1868363}
\showDOI{\tempurl}


\bibitem[\protect\citeauthoryear{Moreno-Le{\'o}n and Robles}{Moreno-Le{\'o}n
  and Robles}{2014}]%
        {moreno2014}
\bibfield{author}{\bibinfo{person}{Jes{\'u}s Moreno-Le{\'o}n} {and}
  \bibinfo{person}{Gregorio Robles}.} \bibinfo{year}{2014}\natexlab{}.
\newblock \showarticletitle{Automatic detection of bad programming habits in
  scratch: A preliminary study}. In \bibinfo{booktitle}{\emph{2014 IEEE
  Frontiers in Education Conference (FIE) Proceedings}}. \bibinfo{pages}{1--4}.
\newblock
\showISSN{2377-634X}
\urldef\tempurl%
\url{https://doi.org/10.1109/FIE.2014.7044055}
\showDOI{\tempurl}


\bibitem[\protect\citeauthoryear{Sorva}{Sorva}{2018}]%
        {sorva2018}
\bibfield{author}{\bibinfo{person}{Juha Sorva}.}
  \bibinfo{year}{2018}\natexlab{}.
\newblock \bibinfo{booktitle}{\emph{Misconceptions and the Beginner
  Programmer}}.
\newblock
\showISBNx{9781350057111}


\bibitem[\protect\citeauthoryear{Stahlbauer, Kreis, and Fraser}{Stahlbauer
  et~al\mbox{.}}{2019}]%
        {stahlbauer2019testing}
\bibfield{author}{\bibinfo{person}{Andreas Stahlbauer}, \bibinfo{person}{Marvin
  Kreis}, {and} \bibinfo{person}{Gordon Fraser}.}
  \bibinfo{year}{2019}\natexlab{}.
\newblock \showarticletitle{Testing scratch programs automatically}. In
  \bibinfo{booktitle}{\emph{Proceedings of the 2019 27th ACM Joint Meeting on
  European Software Engineering Conference and Symposium on the Foundations of
  Software Engineering}}. \bibinfo{pages}{165--175}.
\newblock


\bibitem[\protect\citeauthoryear{Swidan, Hermans, and Smit}{Swidan
  et~al\mbox{.}}{2018}]%
        {swidan2018}
\bibfield{author}{\bibinfo{person}{Alaaeddin Swidan}, \bibinfo{person}{Felienne
  Hermans}, {and} \bibinfo{person}{Marileen Smit}.}
  \bibinfo{year}{2018}\natexlab{}.
\newblock \showarticletitle{Programming Misconceptions for School Students}. In
  \bibinfo{booktitle}{\emph{Proceedings of the 2018 ACM Conference on
  International Computing Education Research}} \emph{(\bibinfo{series}{ICER
  ’18})}. \bibinfo{publisher}{Association for Computing Machinery},
  \bibinfo{address}{New York, NY, USA}, \bibinfo{pages}{151–159}.
\newblock
\showISBNx{9781450356282}
\urldef\tempurl%
\url{https://doi.org/10.1145/3230977.3230995}
\showDOI{\tempurl}


\bibitem[\protect\citeauthoryear{Techapalokul and Tilevich}{Techapalokul and
  Tilevich}{2017a}]%
        {techapaloku2017b}
\bibfield{author}{\bibinfo{person}{Peeratham Techapalokul} {and}
  \bibinfo{person}{Eli Tilevich}.} \bibinfo{year}{2017}\natexlab{a}.
\newblock \showarticletitle{Quality Hound — An online code smell analyzer for
  scratch programs}. In \bibinfo{booktitle}{\emph{2017 IEEE Symposium on Visual
  Languages and Human-Centric Computing (VL/HCC)}}. \bibinfo{pages}{337--338}.
\newblock
\showISSN{1943-6106}
\urldef\tempurl%
\url{https://doi.org/10.1109/VLHCC.2017.8103498}
\showDOI{\tempurl}


\bibitem[\protect\citeauthoryear{Techapalokul and Tilevich}{Techapalokul and
  Tilevich}{2017b}]%
        {techapalokul2017a}
\bibfield{author}{\bibinfo{person}{Peeratham Techapalokul} {and}
  \bibinfo{person}{Eli Tilevich}.} \bibinfo{year}{2017}\natexlab{b}.
\newblock \showarticletitle{Understanding Recurring Quality Problems and Their
  Impact on Code Sharing in Block-Based Software}. In
  \bibinfo{booktitle}{\emph{2017 {{IEEE Symposium}} on {{Visual Languages}} and
  {{Human}}-{{Centric Computing}} ({{VL}}/{{HCC}})}} (2017-10).
  \bibinfo{publisher}{{IEEE}}, \bibinfo{pages}{43--51}.
\newblock
\showISBNx{978-1-5386-0443-4}
\urldef\tempurl%
\url{https://doi.org/10.1109/VLHCC.2017.8103449}
\showDOI{\tempurl}


\end{thebibliography}


\end{document}